# Epitaxial Growth and Characterization of AlInN Based Core-Shell Nanowire Light Emitting Diodes Operating in the Ultraviolet Spectrum

**Ravi Teja Velpula [1], Moab Rajan Philip[1], Barsha Jain[1], Hoang Duy Nguyen[2,¥], Renjie Wang[3], and Hieu Pham Trung Nguyen[1,\*]**

[1]*Department of Electrical and Computer Engineering, New Jersey Institute of Technology, 323 Dr Martin Luther King Jr Boulevard, Newark, New Jersey, 07102*

[2]*Institute of Chemical Technology, Vietnam Academy of Science and Technology, 1 Mac Dinh Chi Street, District 1, Ho Chi Minh City 700000 Vietnam*

[3]*Department of Engineering Physics, McMaster University, 1280 Main Street West, Hamilton, Ontario, L8S 4L7 Canada*

*E-mail: [\*]hieu.p.nguyen@njit.edu; [¥]nhduy@iams.vast.vn*

**Abstract:** We report on the demonstration of the first axial AlInN ultraviolet core-shell nanowire light-emitting diodes with highly stable emission in the UV wavelength range. During the epitaxial growth of AlInN layer, an AlInN shell is spontaneously formed, resulted in the reduced nonradiative recombination on nanowire surface. The AlInN nanowires exhibit high internal quantum efficiency of ~ 52% at room temperature for emission at 295nm. The peak emission wavelength can be varied from 290 nm to 355 nm by changing the growth condition. Moreover, significantly strong transverse magnetic (TM) polarized emission is recorded which is ~ 4 times stronger compared to the transverse electric (TE) **polarized light at 295 nm. This study provides alternative approach for the fabrication of new type of high-performance ultraviolet light-emitters.**

## Introduction

Ultraviolet (UV) light-sources have been in great attention due to their wide range of applications. Such UV light-emitters have primarily used in several key applications include remote detection of biological and chemical compound[1], phototherapy[2], water/air/surface purification and disinfection[3,4], cancer detection[5] and fluorescence sensing or Raman spectroscopy[6]. Currently, major UV light-emitting diode (LED) customers are users of UV-A (315-400 nm) and UV-B (280-315 nm) LEDs, representing over 90% of the overall UV LED market[7]. Among these applications, UV curing is the most dynamic and most important market, due to significant advantages offered over traditional technologies, comprising lower cost of ownership, system miniaturization, etc.[7] Several efforts have been made to develop high efficiency deep UV LEDs using AlGaN material. However, the performance of AlGaN based deep UV LEDs has been fundamentally limited by the large dislocation density and the extremely inefficient *p*-type doing, resulted to low efficiency and low output power[7-10]. Moreover, the device light extraction efficiency is further limited by UV light polarization particularly in the spectral range of ~290-355 nm, when the emitted light polarization switches from transverse electric (TE) polarization to transverse magnetic (TM) polarization due to prohibitively high Al-composition and nanowire geometry.[11,12] As such, polarization state switches from TE to TM reducing light extraction efficiency.[13] Due to these barriers, the external quantum efficiency (EQE) of UV LEDs with emission wavelength above 300 nm can reach up to nearly 10%. However, their performance deteriorates drastically with decreasing wavelengths [14]. For instance, the EQE of deep UV LEDs with emission wavelength below 250 nm decreases dramatically from less than 1% to ~ 0.04%, and the extremely low output power which is just a few tens of nW for emission 210 nm[8]. Such power is extremely low for practical application.

Until recently, fundamental and applied research approaches for light-emitters, have essentially focused on the use of InGaN and AlGaN alloys in the active region for near UV [15,16] and UV photonic devices, respectively, while the approach of using different III-nitride UV materials is relatively unexplored. Identifying and developing the potential of alternative UV materials will be critical to make further progress in development of deep UV emitters. In this regard, $Al_xIn_{1-x}N$ alloy has not been widely studied even though it holds great potential application in UV and visible light-emitting devices. For example, AlInN can be grown perfectly lattice-matched to GaN for an indium content closes to 17–18% [17,18], the large refractive index contrast and high bandgap with respect to GaN make these a highly promising candidate for UV light-emitters compared to AlGaN alloy. Recent studies shown that AlInN offers a large optical gain for deep UV LEDs[19]. Free of defects and quantum-confine Stark effect were achieved for *m*-plan GaN/AlInN multi-quantum wells in core-shell nanowires for UV emitters[20]. Using k.p perturbation theory, Fu et al. reported that AlInN compounds can be grown on both GaN and AlN templates, while AlGaN is detrimental to be grown on GaN templates[21]. AlInN offers wider windows of optimal alloy composition for UV emission compared to AlGaN, especially to deeper UV emission[21]. AlInN has many advantages and is a great of interest that can replace AlGaN or InGaN in several photonic and electronic devices. For instance, lattice matched GaN/AlInN superlattices have been chosen for the near-infrared based on intersubband transitions [22], high reflectivity short wavelength distributed Bragg reflectors (DBRs) [23], high quality factor microcavities for vertical cavity surface emitting laser structures [24], and the realization of high performance high electron mobility transistors (HEMTs) [25].

Albeit holding tremendous advantages, AlInN semiconductor research is highly limited due to the immature epitaxial growth of high quality AlInN. Molecular beam epitaxy (MBE) growth of group-III nitrides under metal-rich conditions usually provides smooth surface morphologies at low growth temperatures. However, under nitrogen-rich growth at these low temperatures results in rough surfaces [26]. The main growth issue for AlInN by MBE is composition inhomogeneity which is commonly presented in AlInN layer [27]. It is suggested that phase separation is attributed to random compositional fluctuations during the early stages of growth, possibly associated with misfit-strain relaxation [27,28]. The difficulties in epitaxial growth of AlInN have resulted from extremely large differences in optimal growth temperatures for InN (~ 450°C) and AlN (~ 800°C) [29]. Additionally, inefficient *p*-type doping in AlInN also strongly affects the electrical properties of the related devices. Such difficulties result in the low crystalline quality and low device performances. The MBE growth under nitrogen rich condition offers an effective approach to eliminate the composition inhomogeneity in the AlInN, reported by Speck et al. [30,31]. By decreasing Al flux and growth of AlInN under N-rich conditions, homogenous AlInN layers with high In content could be achieved [30,31]. Therefore, the nitrogen rich grown nanowires seem to be the best option offering homogenous AlInN structures with nearly-free of dislocation at high In content. However, to our best knowledge, axial nanowire-based AlInN semiconductor grown by MBE has not been reported even though nanowire structures offers several advantages. Nanowire structures offer significant attributes, for instance, significantly improved light output power due to drastically reduced dislocations and polarization fields [32,33]. As intensively reported recently, III-nitride nanowire based LEDs with exceptional performance have been successfully achieved on Si substrates [32,34,35]. More recently, it has been shown that the formation energy for substitutional doping in the near surface region of nanowires can be significantly reduced, together with the nearly-free of dislocation can lead to the enhanced surface doping and conductivity in nanowire LEDs for high-efficiency emission [36].

In this context, we have performed a detailed investigation of the epitaxial growth, structural and optical properties of catalyst-free $Al_xIn_{1-x}N$/GaN nanowires grown on Si (111) substrate by plasma-assisted molecular beam epitaxy (PAMBE). We have also further demonstrated the first axial AlInN core-shell nanowire UV LED heterostructures operating in the UV-A and UV-B bands.

An AlInN shell is spontaneously formed during the growth of AlInN epi-layer which can lead to drastically reduced nonradiative surface recombination. By controlling the Al compositions in the AlInN active region, the emission wavelength can be varied from 290 nm to 355 nm. The AlInN UV nanowires with emission wavelength at 295nm exhibit high internal quantum efficiency (IQE) of ~52% at room temperature. Moreover, the UV LED device exhibits strong UV light emission with highly stable peak emission at 295nm. The polarized optical properties of AlInN nanowire LEDs were also investigated. It is suggested that the UV light from AlInN nanowire LEDs is mainly TM polarized with emission about 4 times stronger than that of TE light.

## Results

### Structural characterizations

In this study, AlInN nanowire light-emitters were grown on $n$-Si (111) substrates by a Veeco Gen II MBE system equipped with a radio-frequency plasma-assisted nitrogen source. Illustrated in Figure 1(a), GaN nanowire template was first grown on Si substrate to facilitate the formation of AlInN segment. Subsequently, detailed study of epitaxial growth of AlInN nanowires on GaN nanowire templates was performed to acquire the optimal growth condition for AlInN nanowire LEDs. Structural properties of AlInN nanowires were characterized by scanning transmission electron microscopy (STEM). Figure 1(b) confirms the presence of GaN and AlInN segments. The wire diameter increase from GaN segment to AlInN portion and remains constant at the top of nanowire which is similar to our reported studies on InGaN, and AlGaN nanowires[35,37,38]. Moreover, it is also suggested that a core-shell AlInN/GaN structure is spontaneously formed during the epitaxial growth of AlInN layer. In order to reveal the compositional variation along the nanowire radial direction, energy dispersive X-ray spectrometry (EDXS) analysis was performed along the GaN and AlInN regions which are indicated as lines 1-2 and 3-4 in Figure 1(b), respectively. Shown in Figure 1(c), corresponding to line-scan 1-2, the Ga signal exhibits a maximum in the nanowire center and drops near the sidewalls. In contrast, the Al signal shows clear peaks near the sidewalls and a dip in the core region of the nanowire. The presence of In is also recorded with detected In signal even though In signal is significantly lower compared to Ga and Al signals. Therefore, it is suggested that a unique GaN/AlInN radial core-shell heterostructure were grown. At the top portion of nanowire corresponding to line-scan 3-4, an AlInN shell around the AlInN core is also confirmed by the EDXS line scan, which is illustrated in Figure 1(d). The In signal is well confined in the core region of the nanowire. The Al signal is again maximum at the sidewalls and significantly reduced at the core section of the wire. The thickness of the shell is about 13.6 nm at the nanowire top and reduces gradually to about 8.4 nm at the nanowire bottom. The formation of such core-shell nanowire structures can be well explained by the diffusion-controlled growth mechanism of III-nitride nanowires under nitrogen-rich conditions which was carefully studied in previous studies[39,40]. Moreover, the presence of the shell layer significantly improves the optical properties of the underlying GaN nanowire templates as well as the AlInN core.

We have further demonstrated the AlInN/GaN UV nanowire LEDs on Si substrate utilizing the optimal growth conditions of AlInN nanowire on GaN templates. The device structure is schematically illustrated in Figure 2(a) which includes a ~ 200 nm GaN:Si segment, 100 nm $Al_xIn_{1-x}N$:Si/40 nm $i$-$Al_yIn_{1-y}N$/100 nm $Al_xIn_{1-x}N$:Mg quantum well, and ~10 nm GaN:Mg. The Al and In compositions in the active region can be varied by adjusting the Al/In flux ratios and/or the growth temperatures to control the emission wavelengths of these AlInN UV nanowire LEDs. Illustrated in Figure 2(b), the nanowires are vertically aligned to the substrate and exhibit nearly uniform heights, with diameter at the top nanowire in the range of ~ 90 nm. Such nanowire properties are suitable for device fabrication.

**Optical Characterizations**

Photoluminescence (PL) spectra of AlInN nanowire on GaN templates were measured using a 266 nm diode-pumped solid-state laser as the excitation source. The PL emission was spectrally resolved by a high-resolution spectrometer and detected by a photomultiplier tube. Figure 3(a) shows photoluminescence spectra of different AlInN/GaN nanowire structures which were grown at different growth conditions. It is clearly shown that the peak emissions vary from 290 nm to 355 nm by varying the Al composition in the AlInN layers. In this study, the Al/In BEP ratio was kept constant while the substrate temperature was increased from 670 °C to 720 °C. The nitrogen flow rate was kept at 2.5 sccm. The peak emission is shifted to shorter wavelength when the substrate temperature is increased which is attributed to the increased In adatom desorption at higher growth temperature, resulted in the reduced In composition in the AlInN segment. Shown in Figure 3(a), the peak emission at ~ 365nm is related to the emission from GaN nanowire templates. We also estimated the Al composition in AlInN layer using the room-temperature PL peak energy $E_{Peak}$ via the following equation:

$E_{Peak}(x) \approx E_g(x) = x E_g(AlN) + (1-x) E_g(InN) - b x(1-x)$, where $x$ is the Al composition, Eg is the bandgap energy. In our calculation, $E_g(AlN)$ and $E_g(InN)$ are considered as 6.2 eV [41] and 0.7 eV [41] respectively and b is taken as the bowing parameter. b is chosen to be 3.4 eV [42]. Shown in Figure 3(b), the Al content is estimated to be in the range of 64 %-76 %, corresponding to emission wavelength from 290 nm – 355 nm. At growth temperature of 710 °C, AlInN nanowires with emission wavelength at 295 nm were recorded with strong emission intensity and spectral linewidth of ~ 28 nm. The optical properties of those AlInN/GaN nanowires were further characterized at different temperatures varying from 20 K to 300 K using liquid Helium to estimate their IQE. Presented in Figure 3(c), the AlInN/GaN nanowire exhibits relatively high IQE which is estimated of ~ 52% at room temperature attributed to the strong carrier confinement provided by the AlInN shell and nearly intrinsic AlInN core. The IQE is calculated by comparing the PL intensity at room temperature and 20 K, assuming the IQE at 20 K is near-unity[43].

**Device Performance**

Such vertically aligned AlInN nanowire LEDs are fully compatible with the conventional fabrication process for large area nanowire LED devices. The device fabrication is described in the method section. The UV nanowire LED devices with an areal size of 500×500 µm$^2$ were chosen for characterization. The AlInN LEDs have excellent current-voltage characteristic with low resistance measured at room temperature, shown in Figure 4(a). The leakage current was found to be very small which is about 1 µA at -8 V. Turn on voltage of these UV nanowire LEDs is ~ 5 V which is significantly lower compared to current thin-film AlGaN LEDs at similar wavelength range[44,45] and is also better/comparable to that of currently reported AlGaN UV nanowire LEDs[43,46,47]. Figure 4(b) presents electroluminescence (EL) spectra of the AlInN nanowire LEDs under various injection currents from 5 mA to 100 mA. No obvious shift in the peak wavelength was observed attributed to the negligible quantum-confined Stark effect (QCSE) in the LED structures, further confirming the high crystalline quality of such AlInN nanowire heterostructures. The light emission polarization properties of the AlInN UV nanowire LEDs were also characterized at room temperature. Transverse-magnetic (TM) and transverse-electric (TE) are defined as the electric field parallel (**E//c**) and perpendicular (**E⊥c**) to c-axis, respectively. The measurement was performed at injection current of 10 A/cm$^2$. Illustrated in Figure 5(a), the UV light emission is predominantly TM polarized which is about > 4 times stronger than that of TE polarized emission. This observation agrees well with the simulation results in which the TM polarized emission is more than two orders of magnitude strong than TE polarized light, shown in Figure 5(b). Similar trend of polarization for LEDs using AlGaN at the same UV wavelength regime are also reported by others[46,48]. This result plays important role in the design of surface emitting UV LEDs using AlInN compounds to achieve high light extraction efficiency.

**Discussions**

In addition to study the performance of AlInN UV nanowire LEDs, we have performed detailed simulation comparing the characteristics of AlInN nanowire LEDs with and without the integration of electron blocking layer. It is clearly shown that, in both LED device structures, electron leakage does not exist or is negligible with quite similar electron current density distribution, shown in Figure 6(a). However, the EBL has strong impact on hole injection efficiency. The EBL-free LED has better hole injection efficiency compared to the other. It is observed in the band diagram of the LED with EBL that there is band bending in the valence band at the heterointerface of the EBL and quantum well (See Figure S1(a) in the Supporting Information), resulted in the hole accumulation at the starting portion of the EBL (See Figure S1(b) in the Supporting Information). This phenomenon decreases the hole injection efficiency in the quantum well, as well as leads to higher turn-on voltage of AlInN UV nanowire LEDs with EBL, presented in Figure 6(b). Advantages of AlInN nanowire structures include the intergration of such nanowire UV LED structures on GaN templates as well as simple structure without the employment of an EBL for high device performance. The EBL-free LED structure is particularly important for developing deep UV LEDs since the Al composition almost reaches maximum for deep UV emission (below 240nm). Therefore, the optimal EBL structure is limitted which requires higher bandgap energy to effectively prevent electron overflow. Moreover, the use of EBL will also affect the hole transport, resulted to the reduced hole injection efficiency to the device active region. Further optimization in term of device structure, active region thickness and composition will be performed to achieve high power AlInN deep UV nanowire LEDs.

In summary, we have successfully demonstrated the first AlInN axial nanowire LEDs operating the UV-A and UV-B bands with relatively high IQE of ~ 52% at room temperature. The electron overflow was not observed within these nanowire UV emitters. The devices exhibit stable emission with strong TM polarized light emission. The device performance can be further improved by engineering the device structure, nanowire morphology as well as nanowire diameter and spacing to enhance the light extraction efficiency of such AlInN UV core-shell nanowire LEDs.

## Materials and methods

### Molecular beam epitaxial (MBE) growth

Vertically aligned self-organized AlInN/GaN heterostructures and AlInN core-shell nanowire LEDs were grown on Si(111) substrates by radio frequency plasma-assisted molecular beam epitaxy. The extremely high purity nitrogen generation system was employed to introduce ultrahigh quality nitrogen gas to the RIBER RF-nitrogen plasma cell. This system includes a Delux Nitrogen purifying system with bypass assembly life status indicator, valve control for bypass and purifier and heating control. The oxide on the substrate surface is desorbed *in-situ* at 780 ºC. First, GaN nanowire templates are formed under nitrogen-rich conditions without the use of any external catalyst. The growth conditions of GaN nanowires include a growth temperature of 770 ºC, with a nitrogen flow rate of 1.0 sccm, a forward plasma power of 400 W, and Ga beam equivalent pressure of $6 \times 10^{-8}$ Torr. To achieve UV light emission, self-organized AlInN segments are subsequently grown on top of GaN nanowires. The In composition in the active region can be controlled by varying the In and Al beam flux and/or the substrate temperature. The growth temperature of AlInN active regions is varied to enhance the In incorporation which is controlled in between 670 ºC to 720 ºC. During the epitaxial growth of AlInN segments, the nitrogen flow rate and plasma power are kept of 2.5 sccm and 400 W, respectively.

### Fabrication process

The UV nanowire LED fabrication process includes the following steps. The nanowire LED samples were first cleaned by HCl then HF to remove native oxide on nanowire surface and oxide layer on the backside of Si substrates, respectively. Ti(20 nm)/Au(120 nm) metal layers were then deposited on the backside of Si wafers for *n*-contact. The *p*-metal contact of Ni(10 nm)/Au(10 nm) was deposited on the top of nanowire samples by e-beam evaporation. For enhancing efficient current spreading of this *p*-contact layer, the top portions of nanowires have to be linked together which can conducted by tilting the substrate holder with a certain angle during the deposition. Thick Ni(20 nm)/Au(120 nm) layers were subsequently deposited on top of the device to serve as metal pad. The fabricated devices with Ti/Au and Ni/Au contacts were annealed at ~ 550 ºC for 1 min. Filling material and Indium Tin Oxide (ITO) were not used in this fabrication to eliminate any light-absorption in this UV wavelength range which is different from our visible InGaN/(Al)GaN nanowire LED fabrication. LEDs with chip areas of ~ 500×500 µm$^2$ were fabricated and selected for characterization.

**Transmission electron microscopy (TEM)**

JEOL JEM-2100F equipped with near field emission gun with an accelerating voltage of 200 kV was used to obtain bright-field TEM images. For STEM-and STEM-HAADF imaging same equipment with a cold field emission emitter operated at 200 kV and with an electron beam diameter of approximately 0.1 nm was used.

**Photoluminescence measurement**

A 266 laser (Kimmon Koha) was used as the excitation source for the photoluminescence measurement of the nanowire heterostructure. The photoluminescence was spectrally resolved by a high-resolution spectrometer and detected by a photomultiplier tube (PMT).

**Electrical, electroluminescence and light polarization characterization**

The current-voltage characteristics of AlInN UV core-shell nanowire LEDs were measured Keithley 2400 digital source meter. Electroluminescence emission of the LED devices were collected by an optical fiber and analyzed using an Ocean Optics spectrometer. The light emission polarization set up consists of an optical fiber together with a Glan-Taylor polarizer mounted on a rotating arm. Signal from the lateral surface of AlInN nanowire LEDs was polarization resolved by polarizer and collected and analyzed by Ocean optics spectrometer. The measurement was performed at 20mA CW injection current.

## Acknowledgment

This work is being supported by NJIT, Instrument Usage Seed Grant from Otto H. York Center at NJIT, and Vietnam National Foundation for Science and Technology Development (NAFOSTED) under grant number 103.03-2017.312. Part of the work was performed in the Cornel NanoFabrication Center.

## Conflict of interest

The authors declare that they have no conflict of interest.

## Author contributions

H.P.T.N. designed the experiment. H.P.T.N., R.W., and M. R. P. contributed to the MBE growth and device fabrication. H.D.N., R.T.V., and M.R.P. performed the EL, PL and power measurements. H.D.N. performed TEM studies. R.T.V. and B. J. contributed to the device simulation. H.P.T.N. and R.T.V wrote the manuscript with contributions from other co-authors.

## References

1       Lytvyn, P. M. et al. Polarization effects in graded AlGaN nanolayers revealed by current-sensing and Kelvin probe microscopy. *ACS Applied Materials & Interfaces* **10**, 6755-6763 (2018).

2       Smith, P. K. C. Laser (and LED) therapy is phototherapy. *Photomedicine and Laser Surgery* **23**, 78-80 (2005).

3       Banas, M. A. et al. Final LDRD report: ultraviolet water purification systems for rural environments and mobile applications. (Sandia National Laboratories, 2005).

4       Würtele, M. et al. Application of GaN-based ultraviolet-C light emitting diodes–UV LEDs–for water disinfection. *Water research* **45**, 1481-1489 (2011).

5       Alimova, A. N. et al. Hybrid phosphorescence and fluorescence native spectroscopy for breast cancer detection. *Journal of biomedical optics* **12**, 014004 (2007).

6       Peng, H. et al. Ultraviolet light-emitting diodes operating in the 340 nm wavelength range and application to time-resolved fluorescence spectroscopy. *Applied Physics Letters* **85**, 1436-1438 (2004).

7       Muramoto, Y., Kimura, M. & Nouda, S. Development and future of ultraviolet light-emitting diodes: UV-LED will replace the UV lamp. *Semiconductor Science and Technology* **29**, 084004 (2014).

8       Kneissl, M. et al. Advances in group III-nitride-based deep UV light-emitting diode technology. *Semiconductor Science and Technology* **26**, 014036 (2011).

9       Kneissl, M., Seong, T.-Y., Han, J. & Amano, H. The emergence and prospects of deep-ultraviolet light-emitting diode technologies. *Nature Photonics* **13**, 233 (2019).

10      Khan, A., Balakrishnan, K. & Katona, T. Ultraviolet light-emitting diodes based on group three nitrides. *Nature photonics* **2**, 77 (2008).

11      Kent, T. F. et al. Deep ultraviolet emitting polarization induced nanowire light emitting diodes with $Al_xGa_{(1-x)}N$ active regions. *Nanotechnology* **25**, 455201 (2014).

12      Zhao, S., Djavid, M. & Mi, Z. Surface Emitting, High efficiency near-vacuum ultraviolet light source with aluminum nitride nanowires monolithically grown on silicon. *Nano Letters* **15**, 7006-7009 (2015).

13      Sarwar, A. T., May, B. J., Chisholm, M. F., Duscher, G. J. & Myers, R. C. Ultrathin GaN quantum disk nanowire LEDs with sub-250 nm electroluminescence. *Nanoscale* **8**, 8024-8032 (2016).

14      Monroy, E., Omnès, F. & Calle, F. Wide-bandgap semiconductor ultraviolet photodetectors. *Semiconductor Science and Technology* **18**, R33 (2003).

15      Chiu, C. et al. Improved output power of InGaN-based ultraviolet LEDs using a heavily Si-doped GaN insertion layer technique. *IEEE Journal of Quantum Electronics* **48**, 175-181 (2012).

16      Lin, L. et al. InGaN/GaN ultraviolet LED with a graphene/AZO transparent current spreading layer. *Opt. Mater. Express* **8**, 1818-1826 (2018).

17      Carlin, J.-F. & Ilegems, M. High-quality AlInN for high index contrast Bragg mirrors lattice matched to GaN. *Applied Physics Letters* **83**, 668-670 (2003).

18      Lorenz, K. et al. Anomalous ion channeling in AlInN GaN bilayers: Determination of the strain state. *Physical Review Letters* **97**, 085501 (2006).

19      Tan, C.-K., Sun, W., Borovac, D. & Tansu, N. Large optical gain AlInN-Delta-GaN quantum well for deep ultraviolet emitters. *Scientific Reports* **6**, 22983 (2016).

20      Durand, C. et al. M-plane GaN/InAlN multiple quantum wells in core–shell wire structure for UV emission. *ACS Photonics* **1**, 38-46 (2014).

21      Fu, D. et al. Exploring optimal UV emission windows for AlGaN and AlInN alloys grown on different templates. *physica status solidi (b)* **248**, 2816-2820 (2011).

22      Nicolay, S. et al. Midinfrared intersubband absorption in lattice-matched AlInN⁄GaN multiple quantum wells. *Applied Physics Letters* **87**, 111106 (2005).

23      Dorsaz, J., Carlin, J.-F., Gradecak, S. & Ilegems, M. Progress in AlInN–GaN Bragg reflectors: Application to a microcavity light emitting diode. *Journal of Applied Physics* **97**, 084505 (2005).

24      Feltin, E. et al. Blue lasing at room temperature in an optically pumped lattice-matched AlInN=GaN VCSEL structure. *Electronics Letters* **43**, 924-926 (2007).

25      Gonschorek, M., Carlin, J.-F., Feltin, E., Py, M. A. & Grandjean, N. High electron mobility lattice-matched AlInN⁄GaN field-effect transistor heterostructures. *Applied Physics Letters* **89**, 062106 (2006).

26      Zywietz, T., Neugebauer, J. & Scheffler, M. Adatom diffusion at GaN (0001) and (000Ī) surfaces. *Applied Physics Letters* **73**, 487-489 (1998).

27      Choi, S., Wu, F., Shivaraman, R., Young, E. C. & Speck, J. S. Observation of columnar microstructure in lattice-matched InAlN/GaN grown by plasma assisted molecular beam epitaxy. *Applied Physics Letters* **100**, 232102 (2012).

28      Zhou, L., Smith, D. J., McCartney, M. R., Katzer, D. S. & Storm, D. F. Observation of vertical honeycomb structure in InAlN⁄GaN heterostructures due to lateral phase separation. *Applied Physics Letters* **90**, 081917 (2007).

29      Butté, R. et al. Current status of AlInN layers lattice-matched to GaN for photonics and electronics. *Journal of Physics D: Applied Physics* **40**, 6328 (2007).

30      Stephen, W. K. et al. GaN-based high-electron-mobility transistor structures with homogeneous lattice-matched InAlN barriers grown by plasma-assisted molecular beam epitaxy. *Semiconductor Science and Technology* **29**, 045011 (2014).

31      Kyle, E. C. H., Kaun, S. W., Wu, F., Bonef, B. & Speck, J. S. High indium content homogenous InAlN layers grown by plasma-assisted molecular beam epitaxy. *Journal of Crystal Growth* **454**, 164-172 (2016).

32      Guo, W., Zhang, M., Banerjee, A. & Bhattacharya, P. Catalyst-free InGaN/GaN nanowire light emitting diodes grown on (001) silicon by molecular beam epitaxy. *Nano Letters* **10**, 3355-3359 (2010).

33      Zhao, S., Nguyen, H. P. T., Kibria, M. G. & Mi, Z. III-Nitride nanowire optoelectronics. *Progress in Quantum Electronics* **44**, 14-68 (2015).

34      Hahn, C. et al. Epitaxial growth of InGaN nanowire arrays for light emitting diodes. *Acs Nano* **5**, 3970-3976 (2011).

35      Nguyen, H. P. T. et al. Controlling electron overflow in phosphor-free InGaN/GaN nanowire white light-emitting diodes. *Nano Letters* **12**, 1317–1323 (2012).

36      Zhao, S. et al. Tuning the surface charge properties of epitaxial InN nanowires. *Nano Letters* **12**, 2877-2882 (2012).

37      Nguyen, H. P. T. et al. p-Type modulation doped InGaN/GaN dot-in-a-wire white-light-emitting diodes monolithically grown on Si(111). *Nano Letters* **11**, 1919-1924 (2011).

38      Wang, Q., Nguyen, H. P. T., Cui, K. & Mi, Z. High efficiency ultraviolet emission from $Al_xGa_{(1-x)}N$ core-shell nanowire heterostructures grown on Si (111) by molecular beam epitaxy. *Applied Physics Letters* **101**, 043115-043114 (2012).

39      Nguyen, H. P. T. et al. Breaking the carrier injection bottleneck of phosphor-free nanowire white light-emitting diodes. *Nano Letters* **13**, 5437-5442 (2013).

40      Nguyen, H. P. T. et al. Engineering the carrier dynamics of InGaN nanowire white light-emitting diodes by distributed p-AlGaN electron blocking layers. *Scientific Reports* **5**, 7744 (2015).

41      Wu, J. Q. When group-III nitrides go infrared: New properties and perspectives. *Journal of Applied Physics* **106**, (2009).

42      Piprek, J. Nitride semiconductor devices: principles and simulation. (Berlin: Wiley-vch , 2017).

43      Zhao, S. et al. Aluminum nitride nanowire light emitting diodes: Breaking the fundamental bottleneck of deep ultraviolet light sources. *Scientific Reports* **5**, 8332 (2015).

44      Khan, M. A. et al. 13 mW operation of a 295–310 nm AlGaN UV-B LED with a p-AlGaN transparent contact layer for real world applications. *Journal of Materials Chemistry C* **7**, 143-152 (2019).

45      Priante, D. et al. Highly uniform ultraviolet-A quantum-confined AlGaN nanowire LEDs on metal/silicon with a TaN interlayer. *Opt. Mater. Express* **7**, 4214-4224 (2017).

46      Sadaf, S. M. et al. An AlGaN core–shell tunnel junction nanowire light-emitting diode operating in the ultraviolet-C band. *Nano Letters* **17**, 1212-1218 (2017).

47      Growden, T. A. et al. Near-UV electroluminescence in unipolar-doped, bipolar-tunneling GaN/AlN heterostructures. *Light: Science & Applications* **7**, 17150 (2018).

48      Kolbe, T. et al. Optical polarization characteristics of ultraviolet (In)(Al)GaN multiple quantum well light emitting diodes. *Applied Physics Letters* **97**, 171105 (2010).

## FIGURE CAPTIONS

**Figure 1**     (a) Schematic structure of AlInN nanowire on GaN template. (b) TEM image of AlInN/GaN nanowire in which the presence of core-shell structure is clearly shown. EDXS line scan profile showing the quantitative variation of Ga, In and Al signals along the line 1-2. (c) and variation of In and Al signal along the line 3-4.

**Figure 2**     (a) Schematic illustration of AlInN nanowire LED structure on Si. (b) 45º tilted scanning electron microscopy image of typical AlInN nanowire LED sample showing uniform nanowires on Si.

**Figure 3**     (a) Photoluminescence spectra of AlInN/GaN nanowires. The peak emission varies from 290 nm to 355 nm. (b) Photoluminescence peak wavelength versus estimated Al composition. (c) Temperature dependent photoluminescence intensity of AlInN/GaN nanowires.

**Figure 4**      (a) I-V characteristics of the AlInN UV nanowire LED. The inset shows I-V characteristic of the AlInN UV LED device in semi-log scale. (b) Electroluminescence (EL) spectra of the AlInN UV nanowire LEDs under injection current range of 5-100 mA.

**Figure 5**     (a) TM and TE polarized spectra of the AlInN UV nanowire LED with emission wavelength at 295 nm measured at 10 A/cm$^2$. (b) The simulation result of TM and TE polarized lights of AlInN nanowire LEDs at 295 nm.

**Figure 6**     (a) The simulated electron current density of the AlInN nanowire LEDs with and without using EBL showing similar trend of electron current distribution. (b) The simulated I-V characteristics of AlInN nanowire LEDs with and without EBL.

**Figure 1**

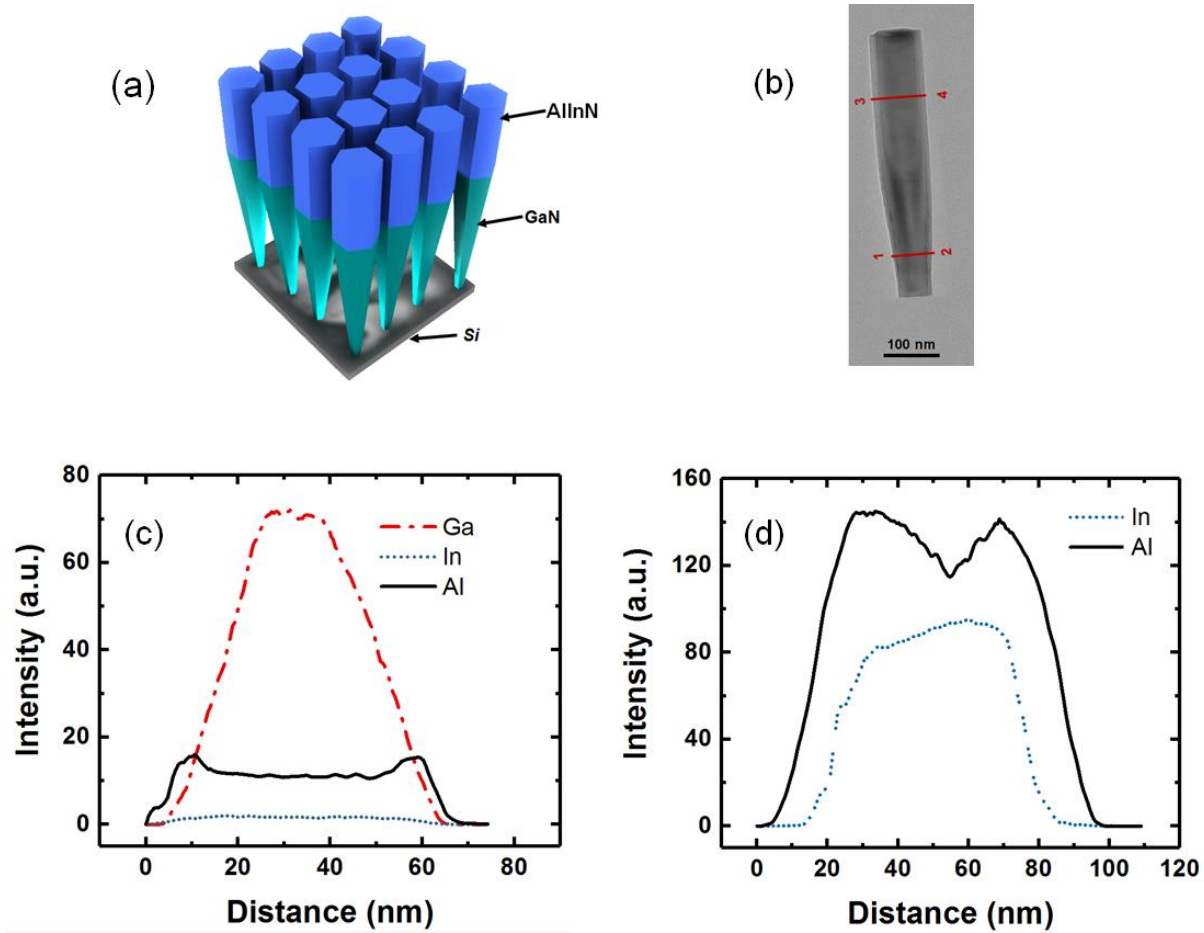

**Figure 2**

(a)

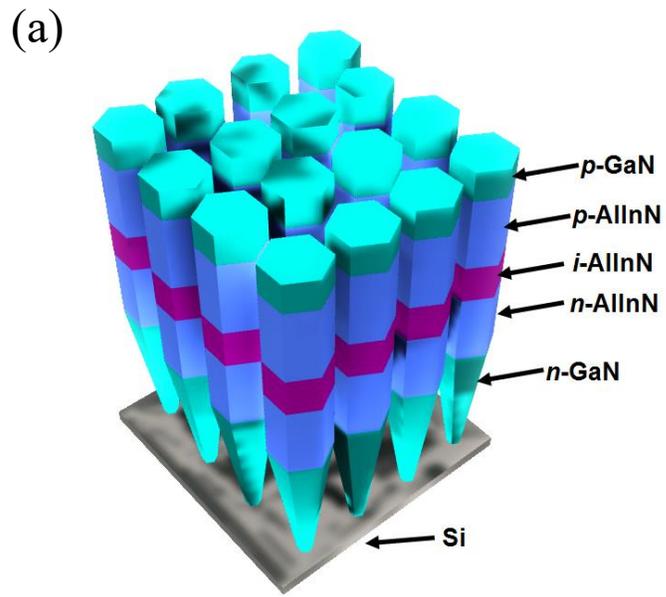

- *p*-GaN
- *p*-AlInN
- *i*-AlInN
- *n*-AlInN
- *n*-GaN

Si

(b)

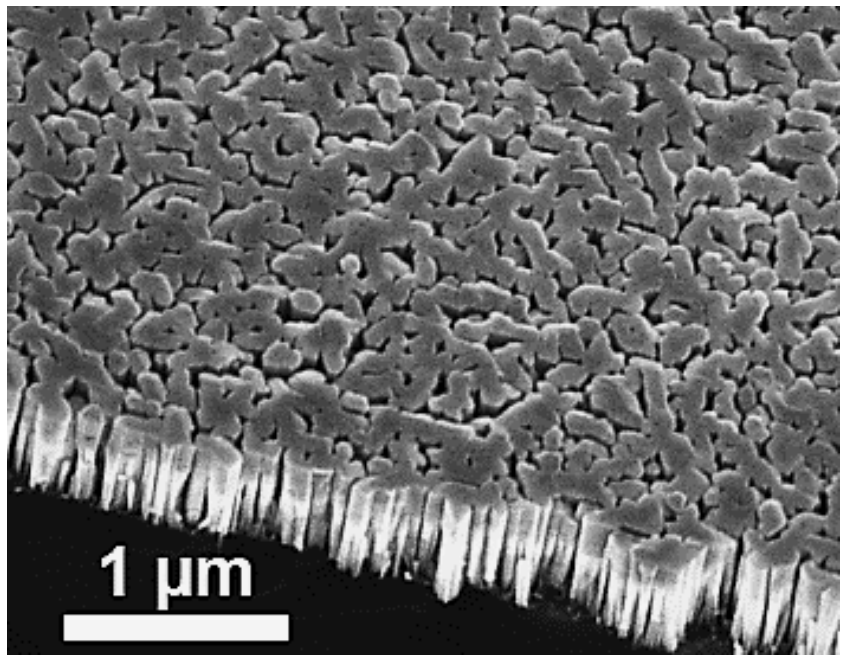

1 µm

**Figure 3**

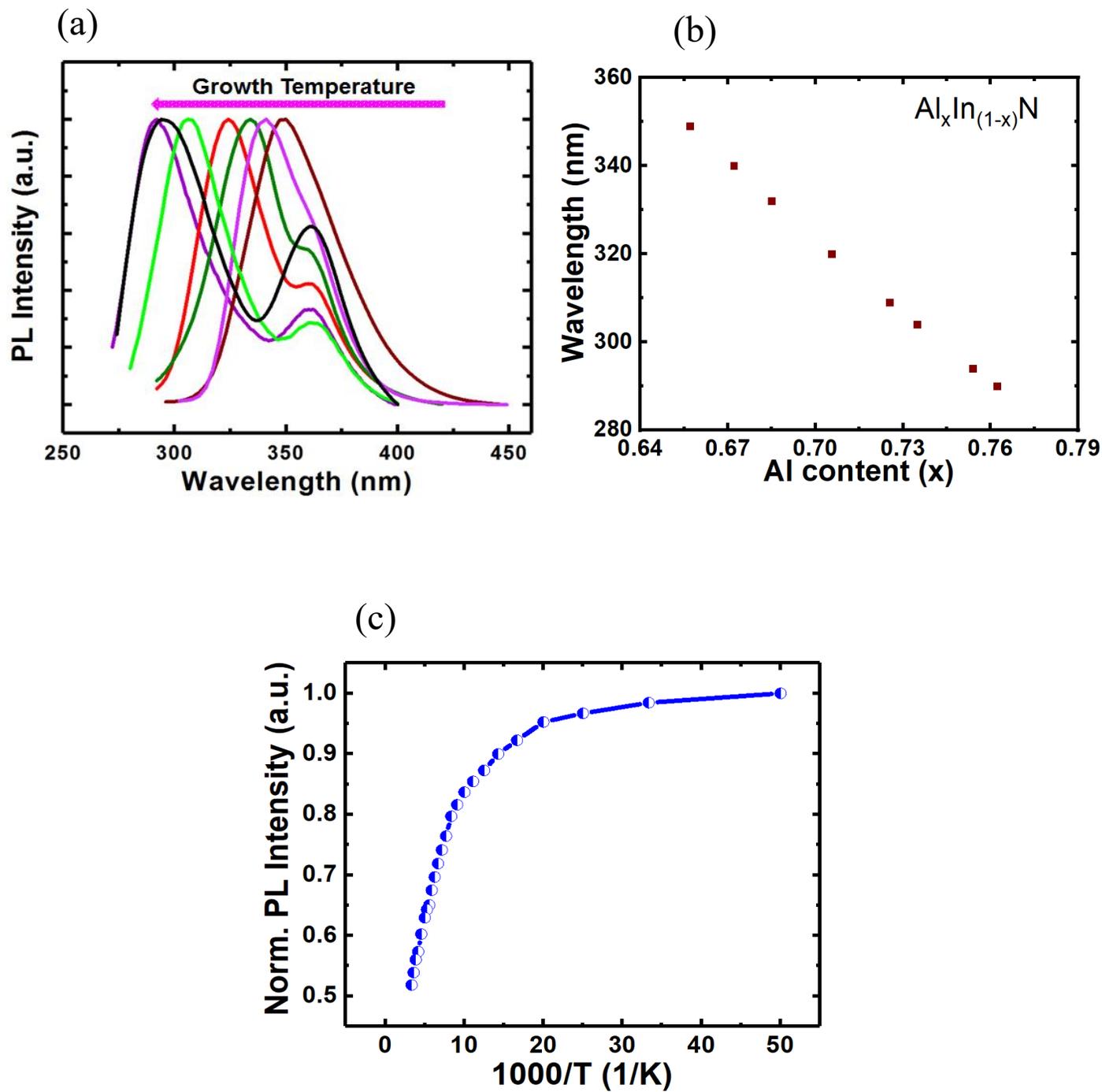

**Figure 4**

(a)

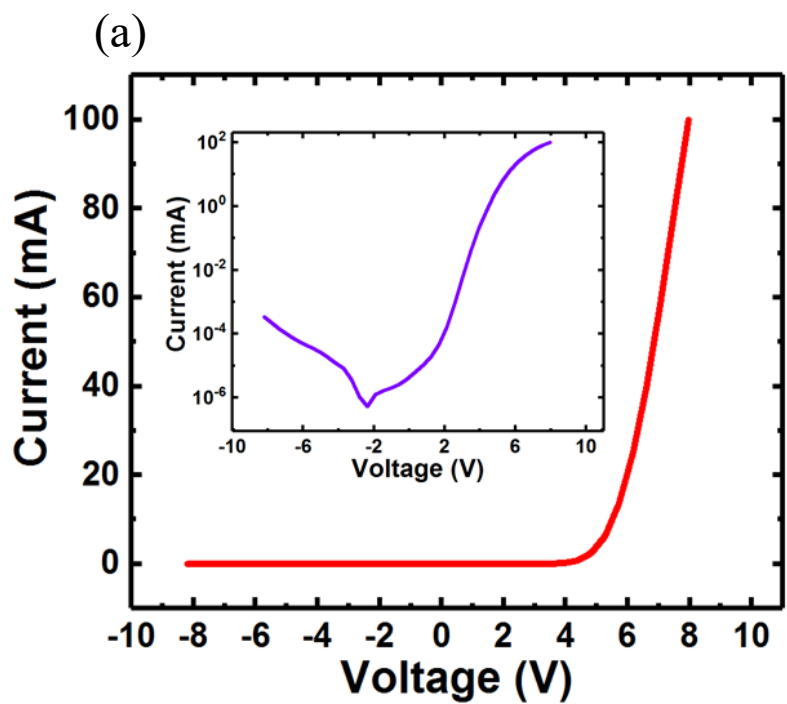

(b)

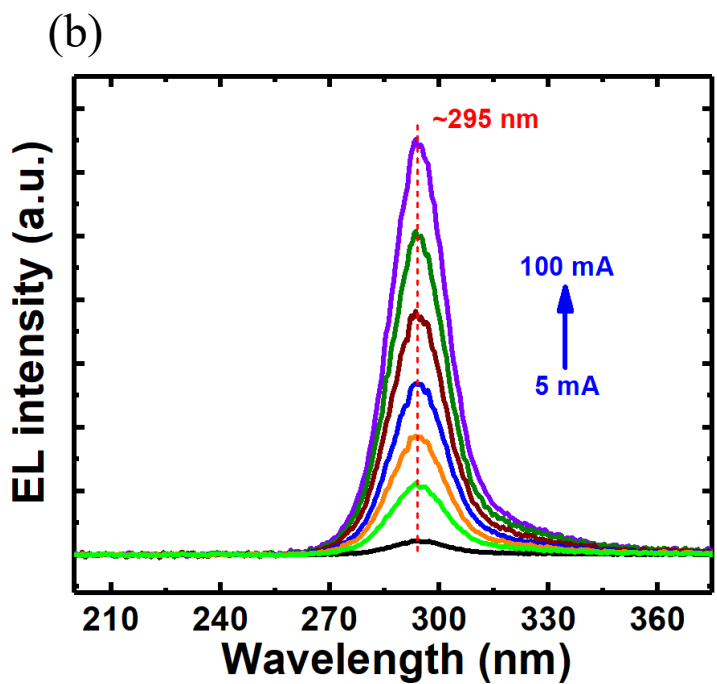

**Figure 5**

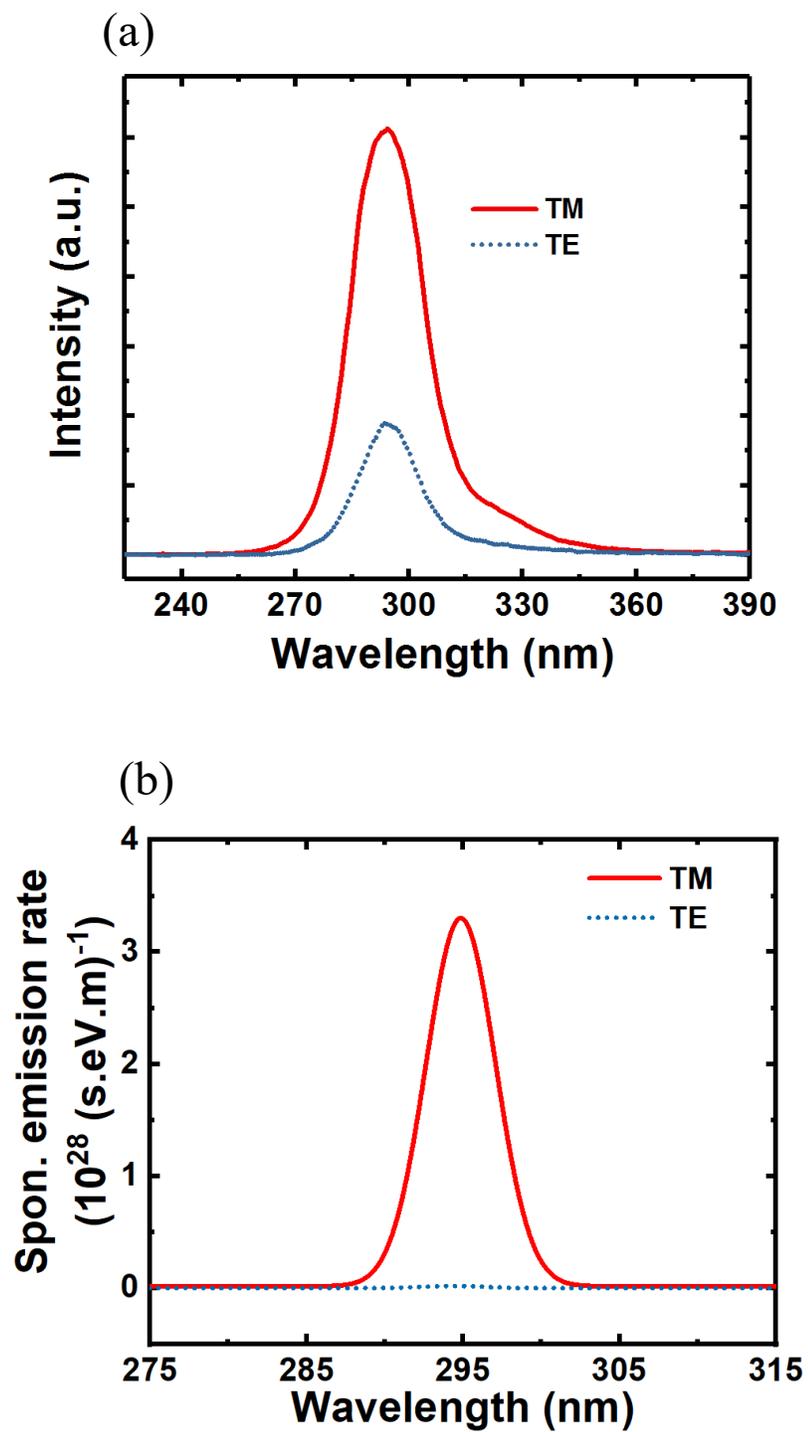

**Figure 6**

(a)

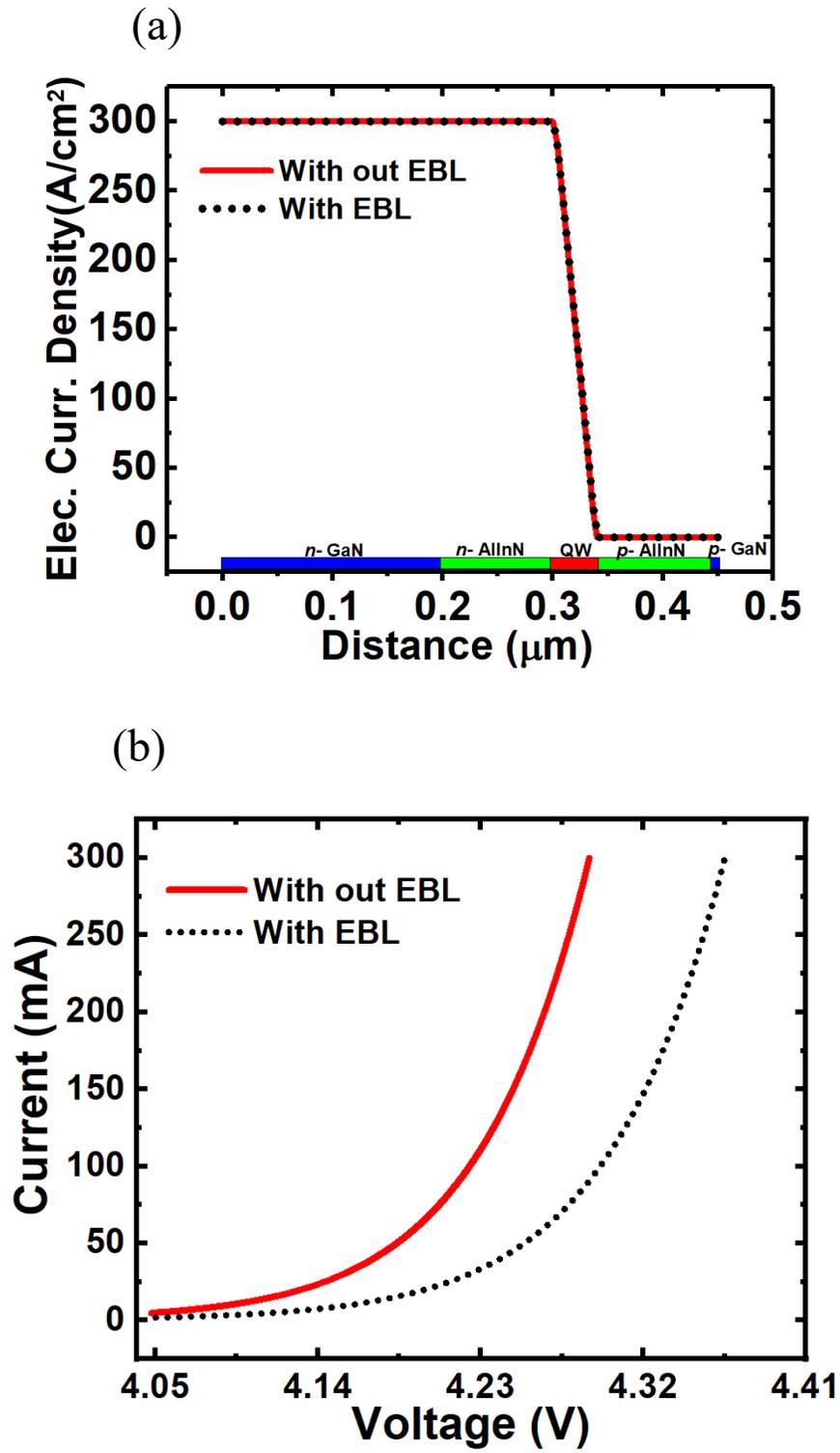

(b)